\documentstyle[preprint,prd,aps]{revtex}

\global\arraycolsep=2pt
\makeatletter
\def\fmslash{\@ifnextchar[{\fmsl@sh}{\fmsl@sh[0mu]}}
\def\fmsl@sh[#1]#2{  \mathchoice
    {\@fmsl@sh\displaystyle{#1}{#2}}    {\@fmsl@sh\textstyle{#1}{#2}}    
{\@fmsl@sh\scriptstyle{#1}{#2}}    {\@fmsl@sh\scriptscriptstyle{#1}{#2}}}
\def\@fmsl@sh#1#2#3{\m@th\ooalign{$\hfil#1\mkern#2/\hfil$\crcr$#1#3$}}
\makeatother

\begin{document}
\draft
\title{Impossibility Criterion for Obtaining Pure Entangled States From Mixed
States By Purifying Protocols}
\author{Ping-Xing Chen\thanks{%
E-mail: pxchen@nudt.edu.cn} and Lin-Mei Liang}
\address{Laboratory of Quantum Communication and Quantum Computation, \\
University of Science and Technology of\\
China, Hefei, 230026, P. R. China \\
and \\
\thanks{%
Corresponding address}Department of Applied Physics, National University of\\
Defense Technology,\\
Changsha, 410073, \\
P. R. China}
\author{Cheng-Zu Li}
\address{Department of Applied Physics,\\
National University of Defense Technology,\\
Changsha, 410073, P. R. China }
\author{Ming-Qiu Huang}
\address{CCAST (World Laboratory) P.O. Box 8730, Beijing, 100080, China\\
and Department of Applied Physics,\\
National University of Defense Technology,\\
Changsha, 410073, P. R. China }
\date{\today}
\maketitle

\begin{abstract}
Purifying noisy entanglement is a protocol which can increase the
entanglement of a mixed state (as a source)at expense of the entanglement of
others(as an ancilla)by collective measurement. A protocol with which one
can get a pure entangled state from a mixed state is defined as purifying
mixed states. We address a basic question: can one get a pure entangled
state from a mixed state? We give a necessary and sufficient condition of
purifying a mixed state by fit local operations and classical communication
and show that for a class of source states and ancilla states in arbitrary
bipartite systems purifying mixed states is impossible by finite rounds of
purifying protocols. For $2\otimes 2$ systems, it is proved that arbitrary
states cannot be purified by individual measurement. The possible
application and meaning of the conclusion are discussed.
\end{abstract}


\thispagestyle{empty}

\newpage \pagenumbering{arabic} 

Quantum information, including quantum computation and quantum communication
has made great advances in recent years\cite{1,2,3,4,5,6,7} .One of
important field is the concentration or purification of entanglement. For a
pure state\cite{2}, n-copy of this state is transformed into m EPR pairs by
Schmidt projection method, and there is no loss of entanglement when n$%
\rightarrow \infty $. For mixed states\cite{3,8,9,10},the entanglement of a
system is increased and another system is destroyed by local unitary
operation and collective measurement\cite{8}. Purifying noisy entanglement
has important applied background in error correcting code\cite{11}, dense
coding\cite{12} and teleportation\cite{13}, etc. There are two different
types of measurement in purifying protocols\cite{8}: individual
measurement(IM)and collective measurement(CM). Suppose Alice and Bob share
an imperfect EPR pair, which is regarded as a source system(SS) of
purification, and an entangled ancilla system(AS) to be destroyed. We note
the source state(SS) as:

\begin{equation}
\rho _s=\sum_ip_i\left| \Psi _i\right\rangle \left\langle \Psi _i\right|
\end{equation}
and the ancilla state(AS) as:

\begin{equation}
\rho _a=\sum_j\lambda _j\left| \Phi _j\right\rangle \left\langle \Phi
_j\right|
\end{equation}
where $\Psi _i$ and $\Phi _j$ are the eigenvectors corresponding to the
nonzero eigenvalues $p_i$ and $\lambda _j$of $\rho _s$ and $\rho _a$,
respectively. Any purifying protocol can be conceived as successive rounds
of local unitary operation, which acts on SS and AS, and measurement on AS
with the help of classical communication. For IM, after a round of purifying
protocols, the state of SS can be written as\cite{6,10}

\begin{equation}
\rho _{sf}=\frac{A_1\otimes B_1\rho _sA_1^{+}\otimes B_1^{+}}{tr(A_1\otimes
B_1\rho _sA_1^{+}\otimes B_1^{+})}  \label{111}
\end{equation}
where $A_1(B_1)$ is an arbitrary operator(in general non-Hermitian)which
acts on the Hilbert space of SS of Alice ( Bob), and necessitate the help of
AS. In general, the noisy entanglement can not be purified by IM\cite{8}.
But any noisy entanglement can be purified with CM. After a round of
purifying protocols with CM, the state of SS can be written as

\begin{equation}
\rho _{sf}=\frac{C_1\rho _sC_1^{+}}{tr(C_1\rho _sC_1^{+})}  \label{112}
\end{equation}
where $C_1$is also an arbitrary operator(in general non-Hermitian) acting on
the whole Hilbert space of SS.

Here we address a question: can one get a maximally entangled state from a
mixed state with nonzero probability by purifying protocol? Because if one
can get an entangled pure state, one can also get a maximally entangled
state by the filter method\cite{9}with non-zero probability, so our question
can be expressed as: can one get a pure entangled state from a mixed state
with non-zero probability by purifying protocol? This is meaningful for one
to set up a noiseless quantum channel which is required in teleportation,
quantum error-correct code and quantum data compression. Although the
earlier work (for example\cite{3} )implies this is impossible for some
states of $2\otimes 2$ systems, they did not answer this question for
general cases. In this paper, first, we introduce the conception of
quasi-separable state(QSS), then exhibit a sufficient and necessary
condition of getting a pure entangled state from a mixed state. It is shown
that if both the state of AS and the state of SS are QSS in every rounds of
purifying protocols, one cannot get an entangled pure state by finite rounds
of purifying protocols for arbitrary bipartite system; if the state of AS or
the state of SS is a QSS, one also cannot get an entangled pure state by
finite rounds of purifying protocols for $2\otimes 2$ systems. Finally, we
discussed what kind of states are QSS.

Before our proof is given, let us give two definitions:

a.{\it \ new-state}(NS): Any mixed state $\rho $ has infinite sets of pure
state decompositions \cite{16}and every decomposition can became another by
the transformation matrices whose columns are orthonormal vectors. For every
decomposition, e.g. Eq(1), if one lets the pure state $\left| \Psi
_i\right\rangle $ unchanged but change the probability $p_i$ of pure state $%
\left| \Psi _i\right\rangle $ in the real numbers realm (0,1), we say one
gets a {\it new-state} of $\rho $.

b. {\it quasi-separable state}(QSS): we say a state $\rho $ is a QSS if one
or many {\it new-state} of $\rho $ is separable.

We now turn to the proof of our result. Any purifying protocol can be
described as following steps: 1). Prepare a source to be purified and an
entangled ancilla to be destroyed between Alice and Bob. 2). Alice and Bob
implement a local unitary operation on their Hilbert space, respectively.
3). Alice and Bob measure the all particles of AS with a set of product
bases of AS, and then AS collapses into one of its bases with definite
probability. By this protocol the entanglement of SS may be increased or the
entropy of SS may be decreased at the expense of the entanglement of AS. One
may ask: why do Alice and Bob measure only the AS and all particles of AS?
Because our aim is the SS, the direct measurement on SS with its bases will
bring the SS into a separable state. If one measure some particles of AS
then one must trace the other particles to get the state of SS. This will be
not fit\cite{15}for one to get a pure entangled state. So the three steps
above is a fit protocol of purifying.

Lemma: For any mixed state of SS and any mixed state of AS, one can distill
a pure entangled state $\Psi $ from SS with nonzero probability by a round
of the purifying protocols above if and only if one can also distill the
pure state $\Psi $ with different nonzero probability from all NS of SS with
the help of all NS of AS.

Proof: \ The sufficient condition is obvious, let us prove necessary one.
Without loss of generality, we suppose $\rho _s$ and $\rho _a$ belong to two
qubits system(the dimension of the Hilbert space is $2\otimes 2$). Note the
bases of $\rho _s$ as $\left| 00\right\rangle ,\left| 01\right\rangle
,\left| 10\right\rangle ,\left| 11\right\rangle $ and $\rho _a$ as $\left|
\uparrow \uparrow \right\rangle ,\left| \uparrow \downarrow \right\rangle
,\left| \downarrow \uparrow \right\rangle ,\left| \downarrow \downarrow
\right\rangle $. First, Alice and Bob use a fit local unitary operation on
their two qubits, respectively. Then, Alice and Bob measure the $\rho _a$
with a set of orthogonal and locally distinguishable product bases \cite{18}%
, e.g. $\left| \uparrow \uparrow \right\rangle ,\left| \uparrow \downarrow
\right\rangle ,\left| \downarrow \uparrow \right\rangle ,\left| \downarrow
\downarrow \right\rangle ($the other locally distinguishable bases can be
changed into this one by local unitary transformation), and then AS
collapses into a vector(suppose it is $\left| \uparrow \uparrow
\right\rangle ).$The local operator $u$ noted as $u^A\otimes u^B$ act on $%
\rho _s\otimes \rho _a$, i.e:

\begin{equation}
u^A\otimes u^B\rho _s\otimes \rho _au^{+A}\otimes u^{+B}=\sum_{ij}p_i\lambda
_ju^A\otimes u^B\left| \Psi _i\right\rangle \left\langle \Psi _i\right|
\left| \Phi _j\right\rangle \left\langle \Phi _j\right| u^{+A}\otimes u^{+B}
\label{3}
\end{equation}
where $\left| \Psi _i\right\rangle $ and $\left| \Phi _i\right\rangle $ are
an arbitrary set of pure state decomposition of $\rho _s$ and $\rho _a$,
respectively.

We note: 
\begin{equation}
u^A\otimes u^B\left| \Psi _i\right\rangle \left| \Phi _j\right\rangle
=\left| \Psi _{ij}^1\right\rangle \left| \uparrow \uparrow \right\rangle
+\left| \Psi _{ij}^2\right\rangle \left| \uparrow \downarrow \right\rangle
+\left| \Psi _{ij}^3\right\rangle \left| \downarrow \uparrow \right\rangle
+\left| \Psi _{ij}^4\right\rangle \left| \downarrow \downarrow \right\rangle
\label{4}
\end{equation}
where $\left| \Psi _{ij}^m\right\rangle (m=1,2,3,4)$ are pure states of SS.
Different $u^A\otimes u^B$ result in the corresponding $\left| \Psi
_{ij}^m\right\rangle $. One can choose a fit $u^A\otimes u^B$ to get a pure
desirous state $\Psi $ by measurement with bases $\left| \uparrow \uparrow
\right\rangle ,\left| \uparrow \downarrow \right\rangle ,\left| \downarrow
\uparrow \right\rangle ,\left| \downarrow \downarrow \right\rangle .$ After $%
u^A\otimes u^B$act on $\rho _s\otimes \rho _a,$ the state of SS and AS are
still a mixed state. To get a pure state $\Psi $ with nonzero probability
from SS, one should let every $\left| \Psi _{ij}^1\right\rangle $ (for all
i,j) be $\Psi $ or some $\left| \Psi _{ij}^1\right\rangle $ do not exist if
AS collapses into a vector $\left| \uparrow \uparrow \right\rangle .$
Obviously all $\left| \Psi _{ij}^1\right\rangle $ are not dependant on $%
p_i,\lambda _j,$ which change only the probability of getting the state $%
\Psi .$ If for definite $p_i$ and $\lambda _j$ one can get a pure entangled
state, one can also get the same state with different probability when one
only changes the $p_i$ and $\lambda _j$ in the realm (0,1). Because $\left|
\Psi _i\right\rangle $ and $\left| \Phi _i\right\rangle $ are an arbitrary
set of pure state decomposition of $\rho _s$ and $\rho _a$, respectively, we
can say if one can get a entangled pure state $\Psi $ one can also get the
same pure state $\Psi $ from all NS of SS with the help of all NS of AS.
This proof method and result can be generated to any dimension system
easily. So the Lemma is proved. The Lemma imply the following result:

Theorem 1: If both $\rho _s$ and $\rho _a$ are QSS, one cannot get a pure
entangled state $\Psi $ with nonzero probability from the SS by a round of
purifying protocols .

Proof: The proof of theorem is very easy. When the states of both $\rho _s$
and $\rho _a$ are QSS, if one can get a pure entangled state from the SS,
according to the Lemma one can also get the same entangled state from a
separable NS of SS with the help of a separable NS of AS with nonzero
probability. This is impossible\cite{17}

Suppose a state $\rho $ is transferred into a state $\rho ^{^{\prime }}$ by
local operation and classical communication(LOCC), it is obviously that if $%
\rho $ is a QSS, $\rho ^{^{\prime }}$ is also a QSS. Because the {\it %
new-states} of $\rho $ is transferred into the{\it \ new-states} of $\rho
^{^{\prime }}$ under the same LOCC. If both $\rho _s$ and $\rho _a$ are QSS
in every round of purifying protocols, one cannot get a pure entangled state 
$\Psi $ with nonzero probability by any finite rounds of purifying protocols.

Now, we discuss what kind of states are QSS. If a state , the Hilbert
space's dimensions of which are $n$ , is full rank, it has n eigenvectors
with the nonzero eigenvalues. These eigenvectors consist of a set bases of
the Hilbert space. A NS of this state we choose is a mixed state of all
these eigenvectors with equal probability 1/n. From Wootters' scheme\cite{4}%
, one can find a unitary transformation by which one can get a set of
product pure state decomposition of this NS, and this set of pure state is
another set of bases of the Hilbert space. So we obtain the theorem2:

Theorem2: All states of full rank are QSS

The result above are fit to arbitrary dimension system, now we pay more
attention on $2\otimes 2$ systems . A mixed state $\rho $ can be decomposed
into\cite{4} 
\begin{equation}
\rho =\sum_{i=1}^l\left| x_i\right\rangle \left\langle x_i\right|
=\sum_{i=1}^m\left| z_i\right\rangle \left\langle z_i\right| ,  \label{6}
\end{equation}
where $\left| x_i\right\rangle ,$ unnormalized, is a complete set of
orthogonal eigenvectors corresponding to the nonzero eigenvalues of $\rho $,
and $\left\langle x_i\right| x_i\rangle $ is equal to the its nonzero
eigenvalues. For a state of $2\otimes 2$ systems, there exist a set of
decomposition $\left| z_i\right\rangle $ of $\rho $ noted by 
\begin{equation}
\left| z_i\right\rangle =\sum_{j=1}^lu_{ij}\left| x_j\right\rangle ,\qquad
i=1,2,\cdots ,m  \label{7}
\end{equation}
where $\left| z_i\right\rangle $ is not necessarily orthogonal, the columns
of transformation $u_{k\times l}$ are orthonormal vectors, and

\begin{equation}
\left\langle z_i\right| \left. \widetilde{z}_j\right\rangle =\lambda
_i^{^{\prime }}\delta _{ij},  \label{8}
\end{equation}
where, $\left| \widetilde{z}_i\right\rangle =\sigma _y\otimes \sigma
_y\left| z_i^{*}\right\rangle $. Let us suppose $\lambda _1^{^{\prime
}}>\lambda _2^{^{\prime }}>\lambda _3^{^{\prime }}>$ $\lambda _4^{^{\prime
}} $ and $\lambda _1^{^{\prime }}-\lambda _2^{^{\prime }}-\lambda
_3^{^{\prime }}-\lambda _4^{^{\prime }}>0,$ namely $\rho $ is inseparable%
\cite{4}. If not all $\lambda _i^{^{\prime }}$(i=2,3,4) being zero, one can
get a separable state by decreasing the probability appearing $\left|
z_1\right\rangle $. So in this cases the state $\rho $ is a QSS. But if $%
\lambda _2^{^{\prime }}=\lambda _3^{^{\prime }}=\lambda _4^{^{\prime }}=0$,
the $\rho $ state may be not a QSS. This can be demonstrated by the
following example:

Suppose the state of SS is

\begin{equation}
\rho _s=p_1\left| \Phi ^{+}\right\rangle \left\langle \Phi ^{+}\right|
+p_2\left| 01\right\rangle \left\langle 01\right|
\end{equation}
and the AS is

\begin{equation}
\rho _a=\lambda _1\left| 11\right\rangle \left\langle 11\right| +\lambda
_2\left| \Psi ^{+}\right\rangle \left\langle \Psi ^{+}\right|
\end{equation}
where $\Phi ^{+}=(\left| 00\right\rangle +\left| 11\right\rangle )/\sqrt{2}%
,\Psi ^{+}=(\left| 01\right\rangle +\left| 10\right\rangle )/\sqrt{2}.$
Alice and Bob perform controlled-NOT operations by regarding the SS as
``source'' and AS as ``target''. Then Alice and Bob measure the AS with its
basis vectors $\left| 00\right\rangle ,\left| 01\right\rangle ,\left|
10\right\rangle ,\left| 11\right\rangle $. Thus they can get a pure state $%
\Phi ^{+}$ of SS and the AS collapses into $\left| 01\right\rangle $ with
nonzero probability. From the symmetry $\rho _s$ in Eq.(10) and $\rho _a$ in
Eq.(11) are QSS or not simultaneously. By the theorem1 one knows $\rho _s$
and $\rho _a$ are not QSS.

Theorem3: If the state of AS in a $2\otimes 2$ system is a QSS, one cannot
get an entangled pure state from arbitrary SS in a $2\otimes 2$ system.

Proof: The Lemma and its proof imply theorem3 is equivalent to that one
cannot get a entangled pure state from arbitrary SS with the aid of
arbitrary product bases for $2\otimes 2$ systems. A mixed state $\rho _s$ of 
$2\otimes 2$ systems can be written as a mixture of a pure entangled state
and a separable state\cite{19} , i.e,

\begin{equation}
\rho _s=\lambda _1\left| \Psi \right\rangle \left\langle \Psi \right| +\rho
_{sep}  \label{12}
\end{equation}
$\rho _s$ surely includes a mixed state $\rho $:

\begin{equation}
\rho =\lambda _1\left| \Psi \right\rangle \left\langle \Psi \right| +\lambda
_2\left| \Phi \right\rangle \left\langle \Phi \right|  \label{13}
\end{equation}
where $\left| \Psi \right\rangle =a\left| 0\right\rangle _A\left|
0\right\rangle _B+b\left| 1\right\rangle _A\left| 1\right\rangle _B,$ $%
\left| \Phi \right\rangle =(a_1\left| 0\right\rangle _A+a_2\left|
1\right\rangle _A)\otimes (b_1\left| 0\right\rangle _B+b_2\left|
1\right\rangle _B),\left| 0\right\rangle _i$and $\left| 1\right\rangle
_i(i=A $ or $B)$ are the bases of Alice's or Bob's system. A state of AS
which is a product bases is noted as:

\begin{equation}
\rho _a=\left| \uparrow \right\rangle _A\left\langle \uparrow \right|
_A\otimes \left| \uparrow \right\rangle _B\left\langle \uparrow \right| _B
\label{14}
\end{equation}

It is obvious that if one cannot get a pure entangled state from $\rho $ in
Eq(\ref{13}), one cannot purify the state $\rho _s$ in Eq(\ref{12})$,$ with
the aid of $\rho _a$ in Eq(\ref{14})$.$ Now we will try to purify the state $%
\rho $ of SS. First, Alice and Bob use a local unitary operation $u_A$ and $%
u_B$ on the whole state of $\rho $ and $\rho _a$ . We note: 
\begin{equation}
u_A\otimes u_B\left| \Psi \right\rangle \left| \uparrow \right\rangle
_A\left| \uparrow \right\rangle _B=au_A(\left| 0\right\rangle _A\left|
\uparrow \right\rangle _A)u_B(\left| 0\right\rangle _B\left| \uparrow
\right\rangle _B)+bu_A(\left| 1\right\rangle _A\left| \uparrow \right\rangle
_A)u_B(\left| 1\right\rangle _B\left| \uparrow \right\rangle _B)
\label{15.1}
\end{equation}

\begin{equation}
u_A\otimes u_B\left| \Phi \right\rangle \left| \uparrow \right\rangle
_A\left| \uparrow \right\rangle _B=(a_1u_A(\left| 0\right\rangle _A\left|
\uparrow \right\rangle _A)+a_2u_A(\left| 1\right\rangle _A\left| \uparrow
\right\rangle _A))\otimes (b_1u_B(\left| 0\right\rangle _B\left| \uparrow
\right\rangle _B)+b_2u_B(\left| 1\right\rangle _B\left| \uparrow
\right\rangle _B))  \label{15.2}
\end{equation}

\begin{equation}
u_A(\left| 0\right\rangle _A\left| \uparrow \right\rangle _A)=\left| \Psi
_A^1\right\rangle \left| \uparrow \right\rangle _A+\left| \Psi
_A^2\right\rangle \left| \downarrow \right\rangle _A  \label{16.1}
\end{equation}

\begin{equation}
u_A(\left| 1\right\rangle _A\left| \uparrow \right\rangle _A)=\left| \Psi
_A^3\right\rangle \left| \uparrow \right\rangle _A+\left| \Psi
_A^4\right\rangle \left| \downarrow \right\rangle _A  \label{16.2}
\end{equation}

\begin{equation}
u_B(\left| 0\right\rangle _B\left| \uparrow \right\rangle _B)=\left| \Psi
_B^1\right\rangle \left| \uparrow \right\rangle _B+\left| \Psi
_B^2\right\rangle \left| \downarrow \right\rangle _B  \label{16.3}
\end{equation}

\begin{equation}
u_B(\left| 1\right\rangle _B\left| \uparrow \right\rangle _B)=\left| \Psi
_B^3\right\rangle \left| \uparrow \right\rangle _B+\left| \Psi
_B^4\right\rangle \left| \downarrow \right\rangle _B  \label{16.4}
\end{equation}
Second, Alice and Bob measure the AS. Without loss of generality, we suppose
the AS collapse into $\left| \uparrow \right\rangle _A\left| \downarrow
\right\rangle _B.$ There are only two possible cases in which one may get a
pure entangled state from AS: 1). one gets the same pure entangled state
from $\left| \Phi \right\rangle \left| \uparrow \right\rangle _A\left|
\uparrow \right\rangle _B$ as from $\left| \Psi \right\rangle \left|
\uparrow \right\rangle _A\left| \uparrow \right\rangle _B,$ but this is
impossible because $\left| \Phi \right\rangle \left| \uparrow \right\rangle
_A\left| \uparrow \right\rangle _B$ is separable. 2). There is no component $%
\left| \uparrow \right\rangle _A\left| \downarrow \right\rangle _B$ in Eq(%
\ref{15.2}), But this is also impossible, because this will lead to the
state $\left| \Psi _A^1\right\rangle $ and $\left| \Psi _A^3\right\rangle $
are same(this is necessary for the component $\left| \uparrow \right\rangle
_A$ to disappear) or $\left| \Psi _B^2\right\rangle $ is same as $\left|
\Psi _B^4\right\rangle $(this is necessary for the component $\left|
\downarrow \right\rangle _B$ to disappear), and then result in Eq(\ref{15.1}%
) is separable. So one has no way to get a pure entangled state in this
case, and theorem3 is proved. Similarly, it is easy to prove that if the
state of SS in a $2\otimes 2$ system is a QSS, one cannot get an entangled
pure state with the aid of arbitrary AS in a $2\otimes 2$ system.

The conclusions above imply: 1). A pure maximally entangled state mixed with
an arbitrarily small amount of the identity cannot be purified by individual
measurement, even though the state is already ''good enough''---having
arbitrarily high fidelity to the actual pure state. Because any pure
entangled state mixed with the identity is a state of full rank. 2). All
mixed states of $2\otimes 2$ systems cannot be purified by individual
measurement. 3). The fact that purifying a state of $2\otimes 2$ systems
needs collective measurement and an ancilla state which is not a QSS may be
useful for one to design purify protocols, because if one choose a QSS as an
ancilla, the efficiency of purifying (increment of entanglement of SS per
round of protocols)will tend to zero\cite{3}. However, if an ancilla is not
QSS, one may get a maximally entangled state with nonzero efficiency.

In summary, we address a new basic question: whether one can purify a mixed
state or not by finite rounds of purifying protocols. We give a necessary
and sufficient condition for purifying any mixed state. A practical
conclusion--- all states of QSS cannot be purified with the aid of QSS, is
obtained from this condition. We discuss completely the cases of $2\otimes 2$
systems and show it is impossible for one to purify a mixed state of $%
2\otimes 2$ systems by individual measurement. Our conclusion may be
meaningful in noiseless quantum channel. Because if one can get a singlet,
no matter how small the probability is, one may set up a noiseless quantum
channel. Furthermore, the fact that one can get a nearly-maximally entangled
mixed state by many rounds purifying protocols but can not get a
nearly-separable pure entangled state by individual measurement for
two-qubits systems not only show a new extreme of human's ability, but also
show a difference between a mixed state and a pure state in a deeper way.

\acknowledgments  We thank Guangcan Guo for his support to this work.


\begin{references}
\bibitem{15}  After the tracing one can not get a pure state.
\end{references}
\end{document}